\documentclass[epj,english]{webofc}
\usepackage[varg]{txfonts}

\woctitle{International Conference on New Frontiers in Physics 2013}

\begin{document}


\title{Exclusive production of $W$ pairs in CMS}

\author{
Da Silveira, Gustavo G.\inst{1}
\fnsep\thanks{\email{gustavo.silveira@cern.ch}}
, for the CMS Collaboration
}

\institute{
Centre for Cosmology, Particle Physics and Phenomenology (CP3),
Universit\'e Catholique de Louvain {UCL}, Belgium
}


\abstract{
We report the results on the search for exclusive production of $W$ pairs in 
the LHC with data collected by the Compact Muon Solenoid detector in proton-proton collisions at 
$\sqrt{s}$~=~7~TeV. The analysis comprises the two-photon production of a $W$ 
pairs, ${pp\to p\,W^{+}W^{-}\,p\to p\,\nu e^{\pm}\nu\mu^{\mp}\,p}$. Two events are 
observed in data for 
$p_{\textrm{\footnotesize  T}}(\ell)>$~4~GeV, $|\eta(\ell)|<$~2.4 and 
$m(\mu^{\pm}e^{\mp})>$~20~GeV, in agreement with the standard model prediction 
of 2.2~$\pm$~0.4 signal events with 0.84~$\pm$~0.15 background events. Moreover, a 
study of the tail of the lepton pair transverse momentum distribution is performed 
to search for an evidence of anomalous quartic gauge couplings in the 
$\gamma\gamma\to W^{+}W^{-}$ vertex. As no events are observed in data, it results in 
a model-independent upper limits for the anomalous quartic gauge couplings 
$a^{W}_{0,C}/\Lambda^{2}$, which are of the order of 10$^{-4}$.
}

\maketitle


\section{Introduction}
\label{sec:intro}

\begin{figure}
  \centering
  \includegraphics[width=.3\textwidth]{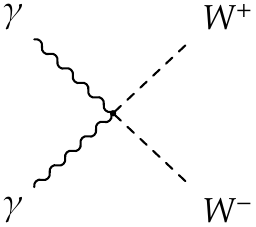}
  \hspace{1em}
  \includegraphics[width=.3\textwidth]{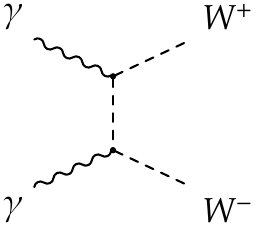}
  \hspace{1em}
  \includegraphics[width=.3\textwidth]{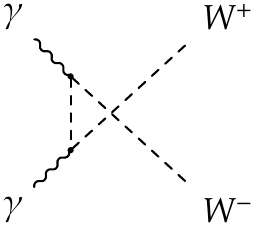}
  \caption{\label{fig:diagrams}
  Diagrams contributing to the $\gamma\gamma\to W^{+}W^{-}$ in the SM: quartic 
  coupling (left), W-boson exchange in the $t$-channel (center), and in the 
  $u$-channel (right).
  }
\end{figure}

Lepton and boson pairs can be produced in elastic and inelastic collisions of 
two colliding protons at high energies by electromagnetic interaction, which 
can be described with the use of electroweak theory. The production 
mechanism by the exchange of two photons in proton-proton collisions, 
$pp \to (\gamma\gamma) \to p+X+p$, has particular signatures that allow its 
identification among other non-exclusive processes, namely, no hadronic activity 
in the final state apart of the particles in central rapidity and 
the forward protons. The latter are scattered at small angles and leave a large 
rapidity gap with respect to the central system. The $\gamma\gamma$ interaction 
provides the possibility to produce lepton pairs $\gamma\gamma\to\ell^{+}\ell^{-}$ 
\cite{Chatrchyan:2011ci}, well-known processes in the framework of electroweak theory with accuracy higher than 
1\%, or boson pairs $\gamma\gamma\to\gamma\gamma,W^{+}W^{-}$ \cite{Abazov:2013opa,Chatrchyan:2013foa}.
In this report, the final state of interest is the two-photon signal 
$\gamma\gamma\to W^{+}W^{-}$, presented in Fig.~\ref{fig:diagrams}, with $W$ bosons decaying into leptons and 
neutrinos.

The elastic (or fully exclusive) events are selected using the silicon tracker 
information in order to reject events with a vertex containing extra tracks other than
the two lepton tracks. On the other hand, the two-photon 
production of $W$ pairs is also accessible in collisions where one or both 
protons dissociate into a hadronic system, labeled here as 
$pp\to p^{*}\,W^{+}W^{-}\,p^{*}\to p^{*}\,e^{\pm}\mu^{\mp}\,p^{*}$.
The search will include these proton dissociative, or quasi-exclusive, events as
part of the $\gamma\gamma\to W^{+}W^{-}$ signal. Considering that the average 
number of simultaneous interaction per bunch crossing (pileup) was 9 during 
the data-taking period of 2011 in the LHC, we look for events with two 
lepton tracks from the same vertex to have a high efficiency in runs at
high luminosities.

Since we search for a $W$ pair decaying into leptons, the backgrounds
will be larger in case of a same-flavor decay like $e^{+}e^{-}$ and 
$\mu^{+}\mu^{-}$, especially due to the Drell-Yan (DY) production. In this 
case, we choose to select events with a $W$ pair decaying into leptons with 
opposite-charge and opposite-flavor, $W^{+}W^{-}\to e^{\pm}\mu^{\mp}$ 
and undetected neutrinos. In view of validating the event selection, a control sample containing 
events from exclusive two-photon production of $\mu^{+}\mu^{-}$ is used, 
since it has small theoretical uncertainties. Also, due to the lack of a Monte Carlo (MC) 
generator accounting for the proton dissociation in $pp\to p\,W^{+}W^{-}\,p$,
we employ this control sample to extract such contribution from data.
The selection of leptons in the final state comprises the lepton pairs with
large transverse momentum $p_{T}(\mu^{\pm}e^{\mp})$ and large invariant mass 
$M(\mu^{\pm}e^{\mp})$ coming from a vertex with no extra tracks apart of the 
two lepton tracks. The result will be then compared to the 
the standard model (SM) expectation for the signal and background events.

Apart of the SM searches, we perform a search for any evidence of new 
physics regarding the quartic coupling, since any deviation from the SM 
prediction can potentially be a signal of new physics. Then, the tail of the 
$p_{T}(\mu^{\pm}e^{\mp})$ distribution are investigated to look for events 
consistent with the predictions for anomalous quartic gauge couplings 
\cite{Belanger:1992qh}.

Therefore, we report the results on the central exclusive production of 
$W$ pairs~\cite{Chatrchyan:2013foa} in proton-proton collisions at $\sqrt{s}$~=~7~TeV obtained 
with the Compact Muon Solenoid (CMS) detector at CERN.

\section{Event selection}
\label{sec:sm-selection}

Regarding the data-taking period of 2011 with collisions at $\sqrt{s}$~=~7~TeV, 
the CMS detector has collected a total of 5.05/fb of data for events in the 
$e^{\pm}\mu^{\mp}$ channel and 5.24/fb in the $\mu^{+}\mu^{-}$ channel.
The online algorithms of the high-level trigger (HLT) select leptons with 
asymmetric thresholds: events with muons with $p_{t}>$~17~GeV 
and electrons/muons with $p_{t}>$~8~GeV and events with 
electrons/muons with $p_{t}>$~17~GeV and muons with $p_{t}>$~8~GeV are selected.


Next, an offline preselection is applied to all leptons in the $e^{\pm}\mu^{\mp}$ sample
and in the control sample:

\begin{itemize}

  \item Reconstructed muon and electron/muon with opposite charge;

  \item Transverse momentum of single leptons: $p_{T}(\ell)>$~20~GeV;

  \item Pseudorapidity, $\eta$, of single leptons: $|\eta(\ell)|<$~2.4;

  \item Number of Extra Tracks: nExtTrk$<$15;

  \item Invariant mass of lepton pairs: $m(\ell^{+}\ell^{\prime -})>$~20~GeV.

\end{itemize}
The kinematic region of interest in this search is defined as events having
no extra tracks associated to 
the $e^{\pm}\mu^{\mp}$ vertex, with pairs having a transverse momentum larger
than 30 GeV in order to reduce the effect of background from 
$\gamma\gamma\to\tau^{+}\tau^{-}$ events. Control plots are produced for
events outside this region to check for the proper modeling of the background
events, as detailed in Sec.~\ref{sec:sm-signal}.


\section{Control sample with $\gamma\gamma\to\mu^{+}\mu^{-}$ events}
\label{sec:sm-control}

Considering the accuracy in the measurement of lepton pairs in 
two-photon processes, we use the $\gamma\gamma\to\mu^{+}\mu^{-}$ sample to validate the selection of lepton 
pairs with high mass. In this case, the selection is divided into two regions 
defined in terms of the $p_{T}$ balance ($|\Delta p_{T}(\mu^{+}\mu^{-})|$) 
and acoplanarity ($1-|\Delta\phi(\mu^{+}\mu^{-})/\pi|$). The first region,
labeled as elastic region, is defined as $|\Delta p_{T}(\mu^{+}\mu^{-})|<$~1~GeV 
an acoplanarity smaller than 0.1~GeV, consistent with the elastic process 
where both protons remain intact in the final state. Besides, the second 
region, related to the proton dissociation, is defined by 
$|\Delta p_{T}(\mu^{+}\mu^{-})|>$~1~GeV and acoplanarity larger than 0.1~GeV. 
The second region is then used to estimate the contribution from the proton 
dissociation in order to be scale the sample of $\gamma\gamma\to W^{+}W^{-}$ 
events.

\begin{figure}
  \centering
  \includegraphics[scale=0.325]{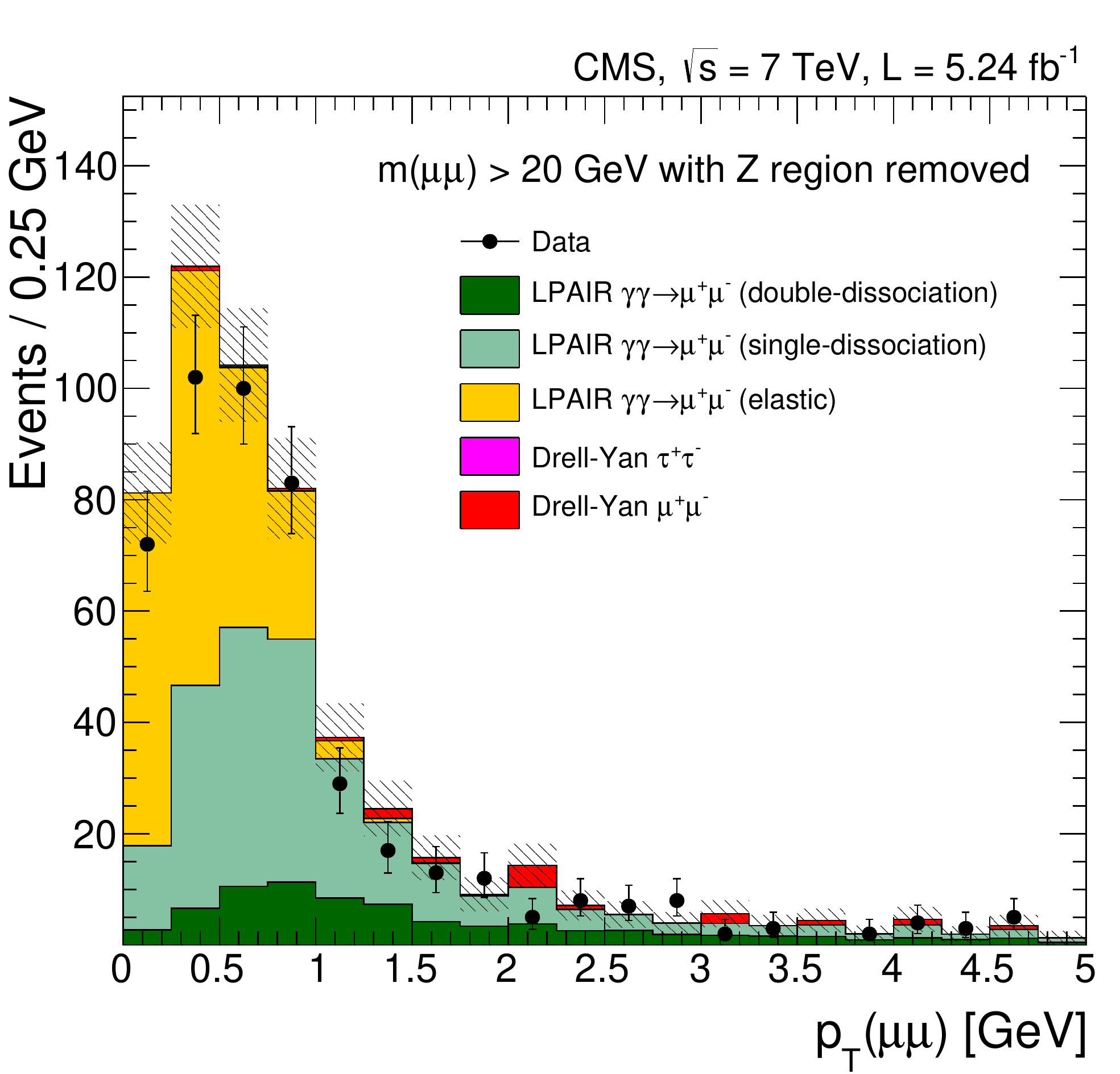}
  \hspace{1em}
  \includegraphics[scale=0.325]{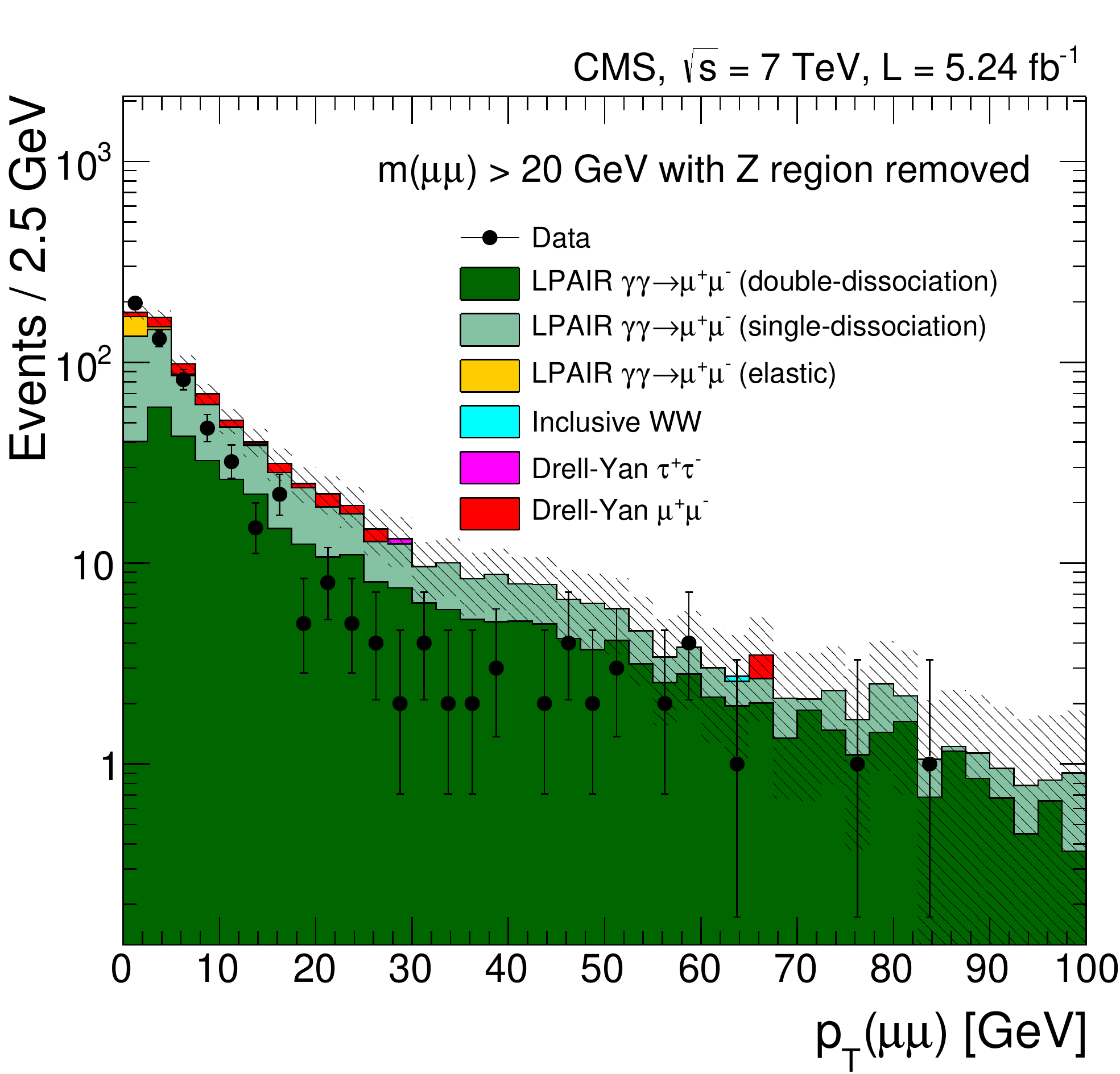}
  \caption{\label{fig:noZpt}
  Transverse momentum distributions for muon pairs with the $Z$-peak region 
  removed (70~${<m(\mu^{+}\mu^{-})<}$~106~GeV) in both elastic (left) and 
  dissociation (right) regions with no extra tracks from the muon pair vertex. 
  The hatched bands indicate the statistical uncertainties in the MC samples.
  }
\end{figure}

To confirm that the MC predictions are properly describing the data
for high-mass lepton pairs. Hence, control plots show that data is well described by 
the theoretical predictions for the two-photon production of $\mu^{+}\mu^{-}$ in the 
region with no extra tracks. Moreover, the $Z$ peak can be used as a 
cross-check considering the large contribution of DY processes, since exclusive and 
photoproduction of the $Z$ bosons are negligible \cite{Goncalves:2007vi,Motyka:2008ac,Cisek:2009hp}. Then, any residual 
background from inclusive DY production can be investigated in the two 
regions defined above. Based on simulation, the $Z$ peak region is defined 
as 70~$<m(\mu^{+}\mu^{-})<$~106~GeV, with the DY $\mu^{+}\mu^{-}$ 
production being dominant. The distributions considering events
outside the $Z$-peak region show that the effects of the DY 
processes are well modeled in this analysis, which allows to perform the search
for $\gamma\gamma\to W^{+}W^{-}$ events outside this region. Figure~\ref{fig:noZpt} presents the transverse 
momentum distribution of the muon pairs in the elastic and dissociation 
regions.

The agreement between MC and data is good in the elastic region, showing that
the {\textsc{lpair}} \cite{Baranov:1991yq,Vermaseren:1982cz} prediction is $\sim$10\% greater than observed in data. 
However, the dissociation region presents a large deficit at large 
$p_{T}(\mu^{+}\mu^{-})$, which correspond to 28\% in comparison to data. This 
region is significantly affected by rescattering corrections when the protons 
dissociate due to the inelastic interaction. This effect can be then estimated 
from data in order to be used to scale the MC predictions for $\gamma\gamma\to W^{+}W^{-}$ events. 
We select the events corresponding to the
exclusive muon pair production with invariant mass higher than 160~GeV and 
divide by the numbers of expected events from theory obtained with {\textsc{lpair}}:
\begin{eqnarray}
  F = \left. \frac{N_{\mu\mu\, \textrm{data}} - N_{DY}}{N_{\textrm{LPAIR}}} \right|_{m(\mu^{+}\mu^{-})>160\,\textrm{GeV}} = 3.23 \pm 0.53.
\end{eqnarray}
Then, this proton-dissociation contribution is included to the predictions of 
exclusive production of $W$ pairs, assuming that the kinematics of the lepton pairs are the same.


\section{Description of data in the signal region}
\label{sec:sm-signal}


The $\gamma\gamma\to W^{+}W^{-}$ signal region is defined for $p_{T}(\mu^{\pm}e^{\mp})>$~30~GeV
and no extra tracks from the vertex. 
As described in Sec.~\ref{sec:sm-control}, the expectation from SM should be scaled to take into 
account the contribution from proton dissociation. The predicted cross section by theory after including the branching ratio for 
leptonic $W$ decays, and it is scaled by the $F$ factor is:
\begin{eqnarray}
  \sigma_{\textrm{th}}(pp\to p^{(*)}W^{+}W^{-}p^{(*)}\to p^{(*)}e^{\pm}\mu^{\mp}p^{(*)}) = 4.0 \pm 0.7\,\textrm{fb}.
\end{eqnarray}

Since the background reasonably well described, the events in the signal region are 
selected within two lepton tracks with no extra tracks from the
vertex and $p_{T}(\mu^{\pm}e^{\mp})>$~30~GeV. Table~\ref{tab:signal} shows the 
number of events passing each stage in the selection chain with the visible 
cross section. We employ the event sample to estimate the 
detector acceptance in the region with $|\eta(\ell)|<$~2.4 and 
$p_{T}(\ell)>$~20~GeV, finding an acceptance in the signal region of 55\%. 

\begin{figure}
  \centering
  \includegraphics[width=.3\textwidth]{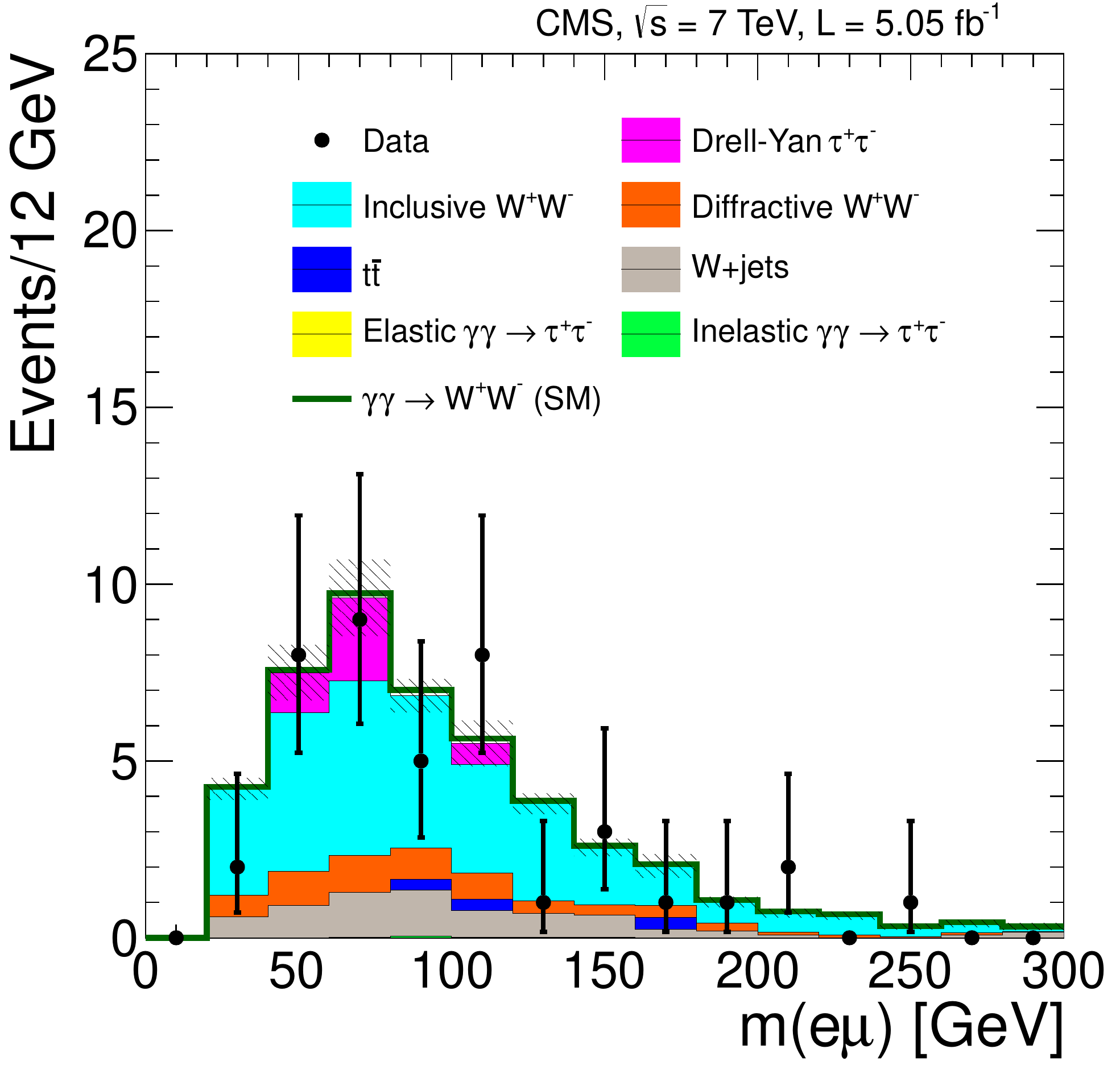}
  \hspace{1em}
  \includegraphics[width=.3\textwidth]{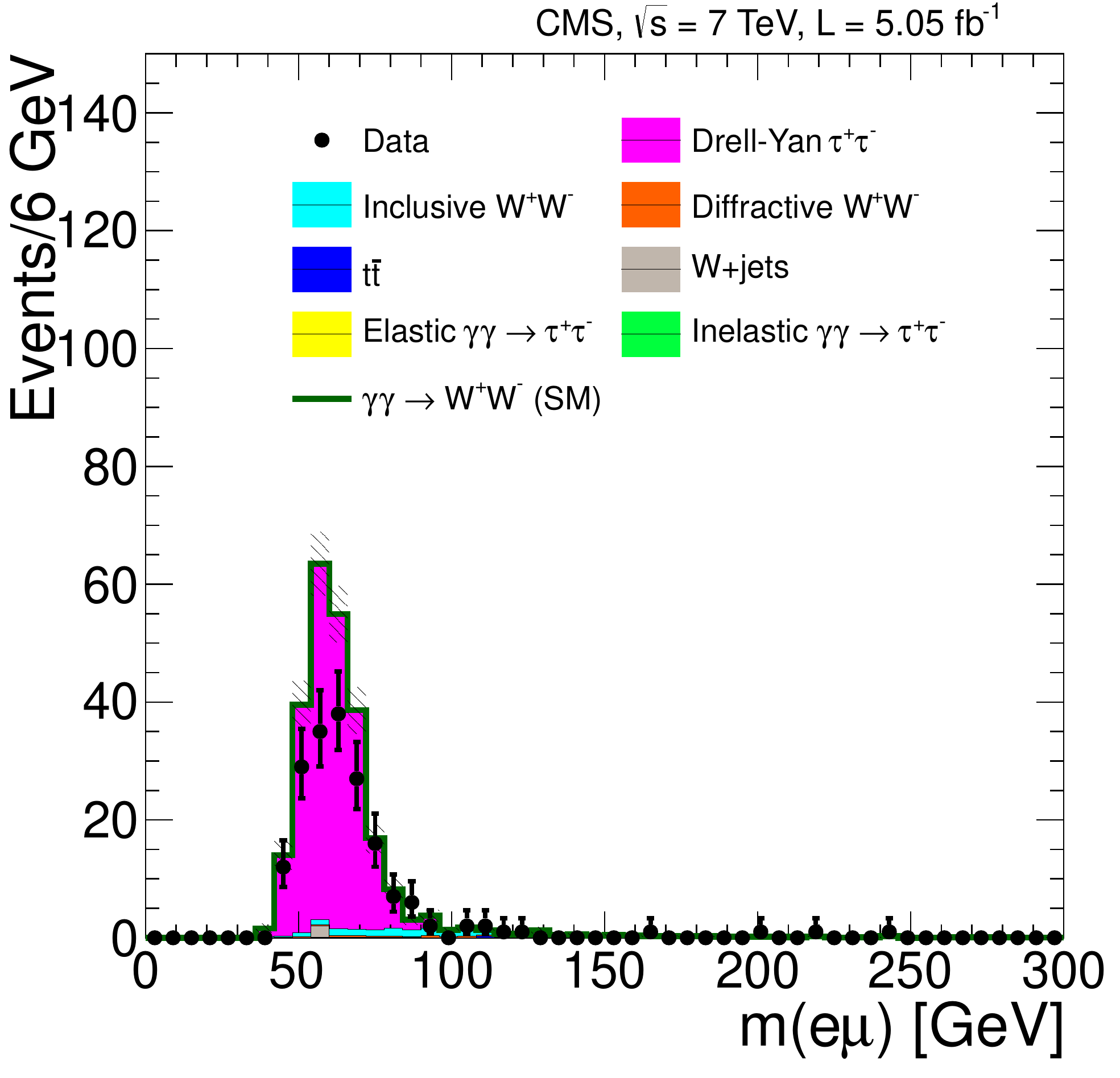}
  \hspace{1em}
  \includegraphics[width=.3\textwidth]{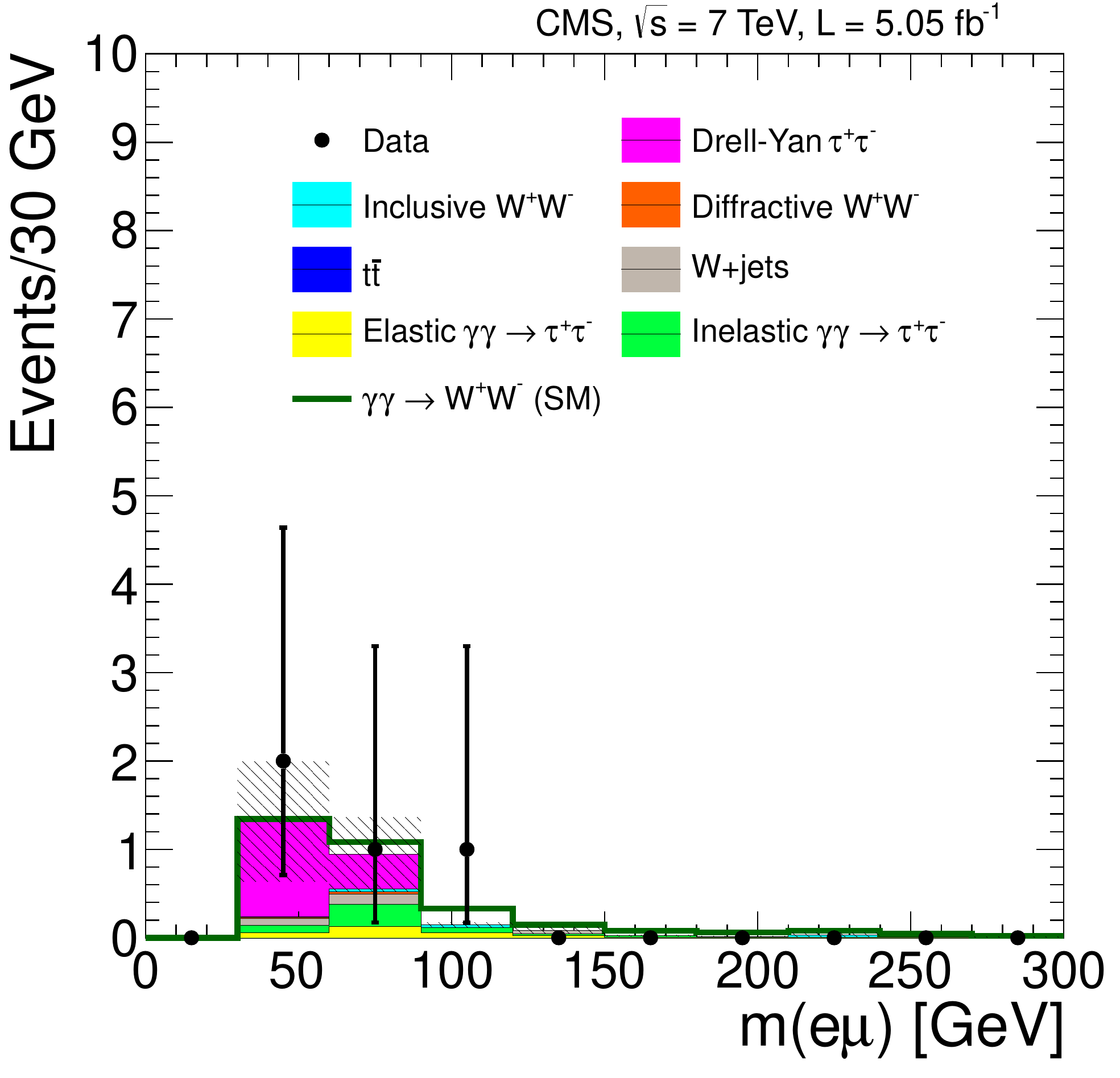}
  \caption{\label{fig:bkgs}
  Invariant mass distribution for each of the three control regions: Region 1 
  (left), Region 2 (center) and Region 3 (right). The shaded bands represent 
  the statistical uncertainty in the MC samples. The signal is stacked to the 
  other histograms.
  }
\end{figure}

Due to the large contribution to the elastic region from background events, we 
defined three control regions for each background processes:

\begin{itemize}

  \item Region 1: Inclusive $W^{+}W^{-}$: $p_{T}(\mu^{\pm}e^{\mp})>$~30~GeV and nExtTrk = 1-6;

  \item Region 2: Inclusive DY $\tau^{+}\tau^{-}$: $p_{T}(\mu^{\pm}e^{\mp})<$~30~GeV and nExtTrk = 1-6;

  \item Region 3: $\gamma\gamma\to\tau^{+}\tau^{-}$: $p_{T}(\mu^{\pm}e^{\mp})<$~30~GeV and nExtTrk = 0.

\end{itemize}
We observe a good overall agreement in all regions, where Table~\ref{tab:regions}
presents the observed backgrounds event yields for the three 
control regions. The theoretical predictions overshoot data in the case of DY 
production of $\tau^{+}\tau^{-}$, however no events survive in the signal region. 
the signal region. Figure~\ref{fig:bkgs} presents the invariant mass distribution 
for each of the three control regions, showing a good description of the data by
the MC predictions. As a result, the background is estimated as 0.84~$\pm$~0.15 events, considering 
the systematic uncertainty on the backgrounds.

\begin{table}
 \caption{\label{tab:regions}
 Background event yields for each of the three control regions.
 }
 \centering  
 \begin{tabular}{ccccc}
  \hline\hline  
  Region & Background & Data & Sum of backgrounds & $\gamma\gamma\rightarrow W^{+}W^{-}$ signal \\  
  \hline
  1 & Inclusive $W^+W^-$                      & 43   & $46.2  \pm 1.7$       & 1.0           \\    
  2 & Inclusive DY $\tau^+\tau^-$      & 182  & $256.7 \pm 10.1$      & 0.3           \\
  3 & $\gamma\gamma\to\tau^+\tau^-$           & 4    & $2.6   \pm 0.8$       & 0.7           \\ 
  \hline\hline  
  \end{tabular}
\end{table} 

\section{Results for the SM searches}
\label{sec:sm-results}

\begin{table}
  \caption{\label{tab:signal}
  Events passing each step in the selection chain, presented with the signal 
  efficiency $\times$ acceptance ($\epsilon\,\times\, A$) and the visible cross section. These events contain 
  reconstructed leptons with opposite charge and flavor, each having 
  $p_{T}(\ell)>$~20~GeV and $|\eta(\ell)|<$~2.4.
  }
  \centering
  \begin{tabular}{lccc}
  \hline\hline   
  Selection step & Signal $\epsilon \times A$ & Visible cross section (fb) & Events in data \\   
  \hline
  Trigger and preselection                         & 28.5\% & 1.1 & 9086 \\
  $m(\mu^{\pm}e^{\mp})>$~20~GeV                    & 28.0\% & 1.1 & 8200 \\
  Muon ID and Electron ID                          & 22.6\% & 0.9 & 1222 \\
  $\mu^{\pm}e^{\mp}$ vertex with zero extra tracks & 13.7\% & 0.6 & 6    \\
  $p_{T}(\mu^{\pm}e^{\mp})>$ 30~GeV                & 10.6\% & 0.4 & 2    \\
  \hline\hline   
  \end{tabular}   
\end{table}

\begin{figure}
  \centering
  \includegraphics[scale=0.325]{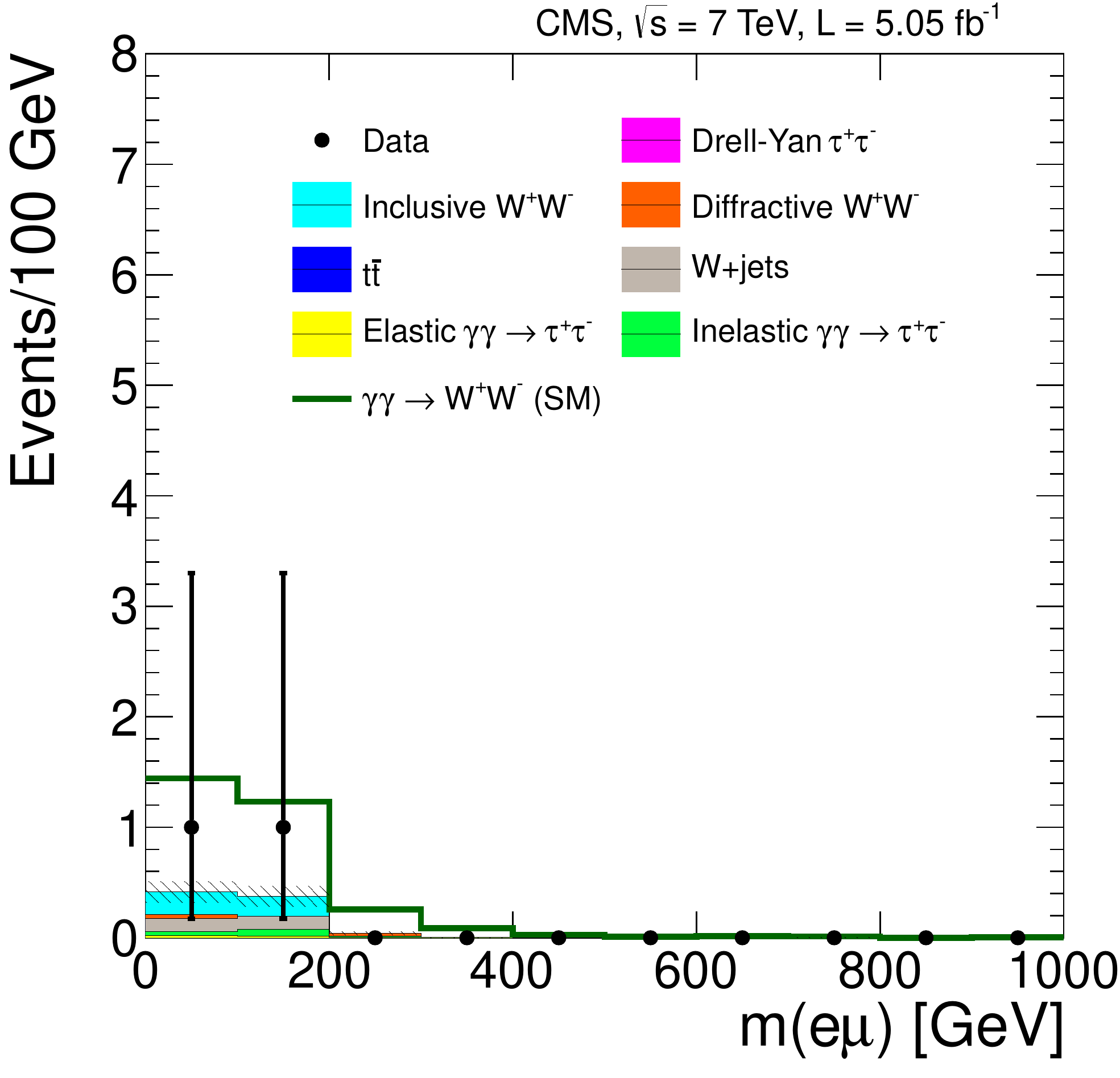}
  \hspace{1em}
  \includegraphics[scale=0.325]{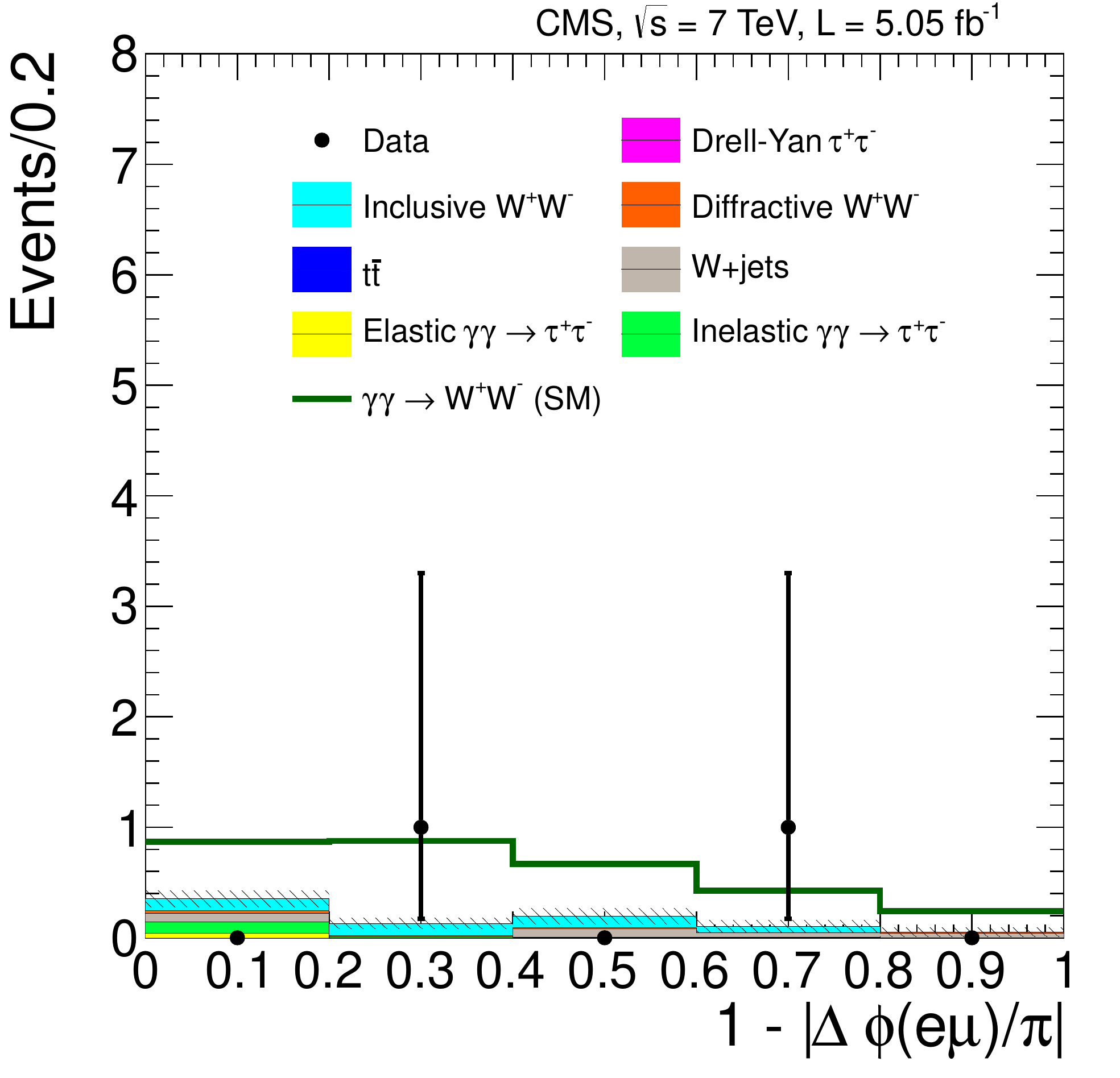}
  \hspace{1em}
  \includegraphics[scale=0.325]{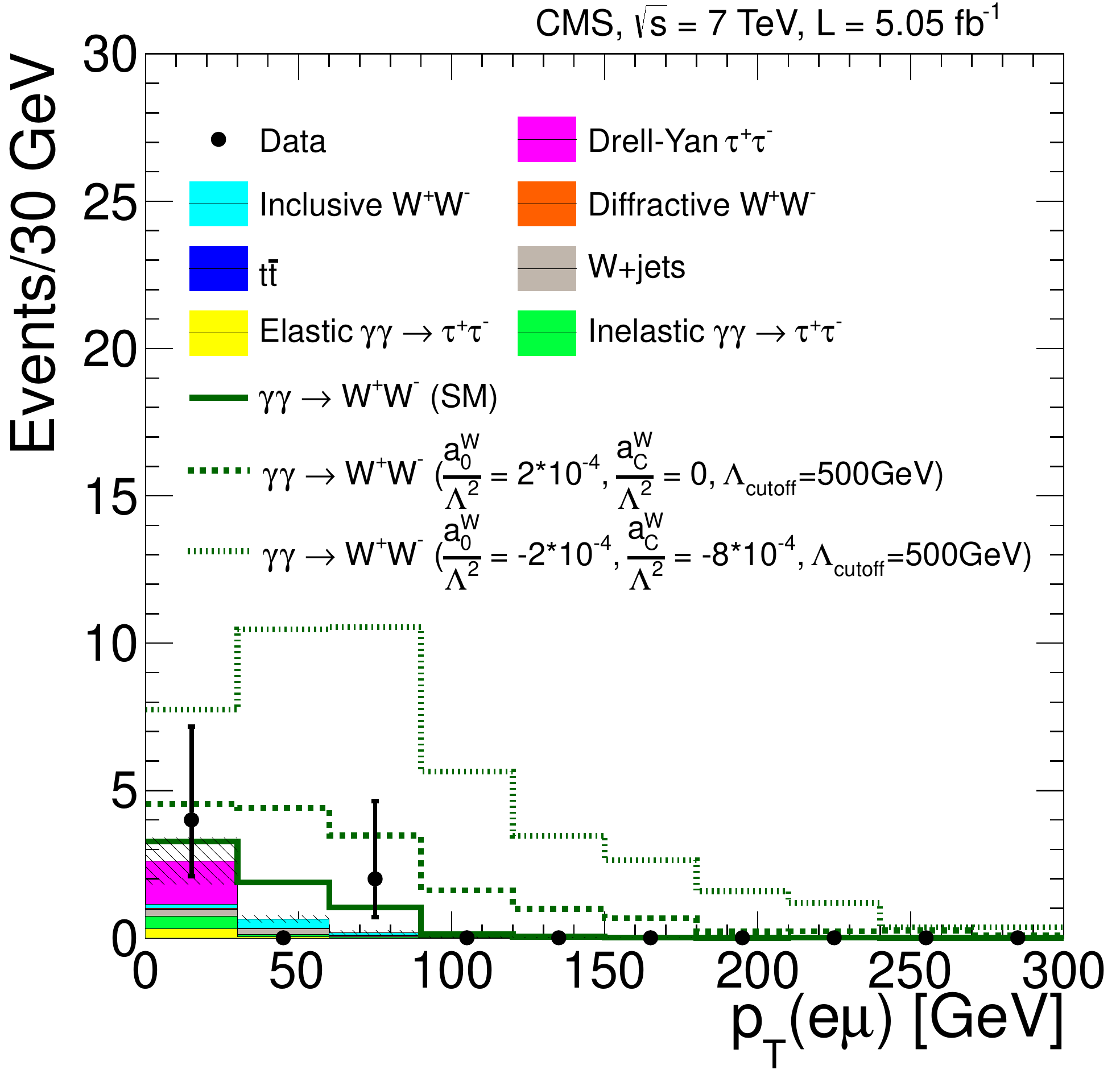}
  \caption{\label{fig:results}
  The $e^{\pm}\mu^{\mp}$ invariant mass (top-left), acoplanarity (top-right) and transverse momentum
  (bottom) distributions in the signal region with no extra tracks. Events are selected with 
  $p_{T}(\ell)>$~20~GeV, $|\eta(\ell)|<$~2.4 and $p_{T}(\mu^{\pm}e^{\mp})>$~30~GeV.
  }
\end{figure}

As a result, two events pass all the selection criteria, 
which is in agreement with the theoretical expectation of 2.2~$\pm$~0.4 of signal events 
and 0.84~$\pm$~0.15 background events. Based on the estimated efficiency $\times$
acceptance ($\epsilon\,\times\, A$) and luminosity ({\cal{L}}), the best fit signal cross section
times branching ratio is:
\begin{eqnarray}
  \sigma(pp\to p^{(*)}W^{+}W^{-}p^{(*)}\to p^{(*)}e^{\pm}\mu^{\mp}p^{(*)}) = 2.2^{+3.3}_{-2.0}\,\textrm{fb},
\end{eqnarray}
with $p_{T}(\ell)>$~20~GeV and $|\eta(\ell)|<$~2.4 with no extra tracks.
Applying the Feldman-Cousins method \cite{Feldman:1997qc}, we estimate the observed upper 
limit in 2.6 times the expect SM yield at 95\% confidence level (CL), which results in the limit on the cross section as:
\begin{eqnarray}
  \sigma(pp\to p^{(*)}W^{+}W^{-}p^{(*)}\to p^{(*)}e^{\pm}\mu^{\mp}p^{(*)}) < 10.6\,\textrm{fb}.
\end{eqnarray}
The invariant mass, the acoplanarity and the transverse momentum distributions are presented in Fig.~\ref{fig:results} with the 
two events in the signal region observed in data.

\section{Beyond SM}
\label{sec:bsm}


It is possible to investigate effects arising from new physics in case that any deviation is 
observed from the SM expectation in the $\gamma\gamma\to W^{+}W^{-}$ vertex. Thus, there are models available in the literature that 
account for anomalous quartic gauge couplings via an effective
Lagrangian \cite{Belanger:1992qh}, which includes the $a^{W,Z}_{0,C}$ anomalous parameters.
Since we expect not enough 
sensitivity to study anomalous triple gauge couplings \cite{Chapon:2009hh}, we focus 
the current study in the anomalous quartic gauge couplings (aQGC), for which there exist limits obtained at LEP 
\cite{Achard:2001eg,Achard:2002iz,Abdallah:2003xn,Abbiendi:1999aa} and at the Tevatron \cite{Abazov:2013opa}, which report anomalous parameters 
of the order of 10$^{-2}$ and 10$^{-3}$, respectively. Moreover, new results from CMS on the 
tri-boson production\cite{CMS:2013kea}, $W\to Z\gamma W$, are reported with limits on the anomalous quartic gauge couplings, 
which shows similar results as the ones found in the exclusive production of $W$ pairs.

Regarding the effective Lagrangian for the aQGC, the resulting cross section 
rises with center-of-mass energy, consequently violating unitarity at 
higher energies. An alternative to tame this rise is apply a dipole form factor 
to each of the anomalous parameters as follows:
\begin{displaymath}
  a^{W}_{0,C}(W^{2}_{\gamma\gamma}) = \frac{a^{W}_{0,C}}{\left( 1+\frac{W^{2}_{\gamma\gamma}}{\Lambda^{2}} \right)^{p}},
\end{displaymath}
with $p=2$ for the dipole form factor and $W_{\gamma\gamma}$ is the center-of-mass 
energy of the $\gamma\gamma$ system. The $\Lambda$ parameter is related to the 
energy scale for new physics and play an important role to 
regulate the cross section. We report results regarding both possibilities of
assuming a scale $\Lambda_{\textrm{cutoff}}=$~500~GeV and no form factor.


\section{Results for the searches beyond SM}
\label{sec:bsm-results}


In order to look for signals of aQGC, we study 
the tail of the $p_{T}(\mu^{\pm}e^{\mp})$ distribution for events with no extra tracks.
For this purpose, we produce event samples with different anomalous parameters by including 
the effective Lagrangian to an event generator. 
By studying the generator-level distributions for $p_{T}(\mu^{\pm}e^{\mp})$ 
for various anomalous parameters, we observe that the SM contribution can 
be neglected for $p_{T}(\mu^{\pm}e^{\mp})>$~100~GeV, where the anomalous 
contribution dominates. Figure~\ref{fig:results} present the 
$p_{T}(\mu^{\pm}e^{\mp})$ distribution with the MC samples for the anomalous couplings. 
No excess is observed in the transverse momentum distribution above 100~GeV.

\begin{figure}
  \centering
  \includegraphics[width=.5\textwidth]{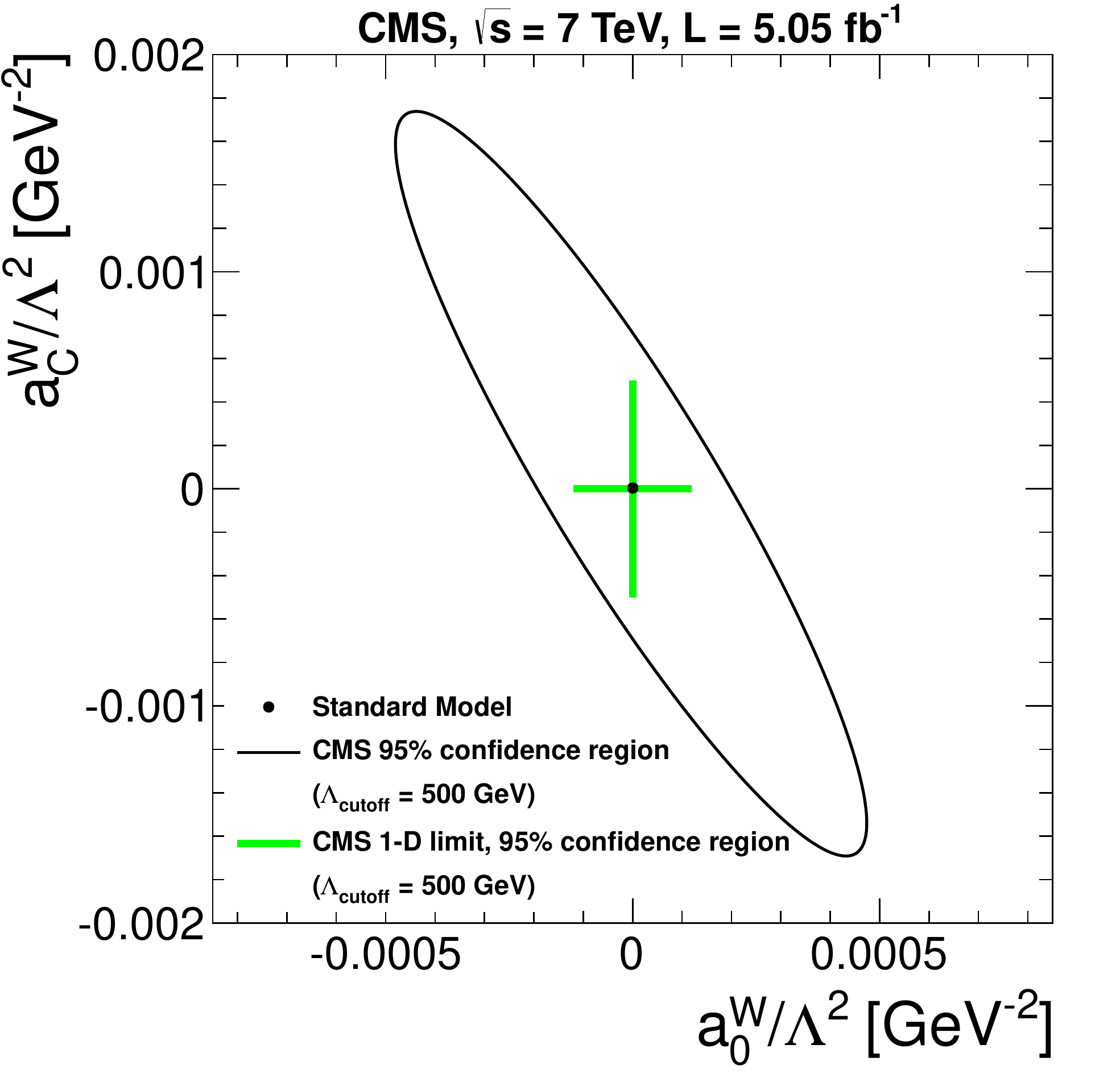}
  \caption{\label{fig:ellipsis}
  Limits on the anomalous couplings $a^{W}_{0}$and $a^{W}_{C}$ for aQGC with 
  $\Lambda_{\textrm{cutoff}}=$~500~GeV. The region outside the ellipsis is 
  excluded with 95\% CL, obtained with $|\eta(\ell)|<$~2.4, $p_{T}(\ell)>$~20~GeV 
  and $p_{T}(e^{\pm}\mu^{\mp})>$~100~GeV.
  }
\end{figure}

In this case, we report an upper limit to the cross section based on the 
Feldman-Cousins method:
\begin{displaymath}
  \sigma(pp\to p^{(*)}W^{+}W^{-}p^{(*)}\to p^{(*)}e^{\pm}\mu^{\mp}p^{(*)}) < 1.9\,\textrm{fb}.
\end{displaymath}
for $|\eta(\ell)|<$~2.4, $p_{T}(\ell)>$~20~GeV and $p_{T}(e^{\pm}\mu^{\mp})>$~100~GeV.
Besides, new limits to the anomalous parameters are reported considering a 
form factor with $\Lambda_{\textrm{cutoff}}=$~500~GeV and without form factors:
\begin{eqnarray*}
  -0.00015 < a^{W}_{0}/\Lambda^{2} < 0.00015~\mathrm{GeV}^{-2}\,\, (a^{W}_{C}/\Lambda^{2} = 0, \Lambda=500~\mathrm{GeV}),\\
  -0.0005 < a^{W}_{C}/\Lambda^{2} < 0.0005~\mathrm{GeV}^{-2}\,\, (a^{W}_{0}/\Lambda^{2} = 0, \Lambda=500~\mathrm{GeV}), \\
  \\
  -4.0 \times 10^{-6} < a^{W}_{0}/\Lambda^{2} < 4.0 \times 10^{-6}~\mathrm{GeV}^{-2}\,\, (a^{W}_{C}/\Lambda^{2} = 0, \mathrm{no~form~factor}),\\
  -1.5 \times 10^{-5} < a^{W}_{C}/\Lambda^{2} < 1.5 \times 10^{-5}~\mathrm{GeV}^{-2} \,\, (a^{W}_{0}/\Lambda^{2} = 0, \mathrm{no~form~factor}).
\end{eqnarray*}
In comparison to the existing limits, the limits presented here are nearly 
two orders of magnitude more stringent than those obtained at LEP and 20 times 
more stringent than the results reported by the Tevatron. Figure~\ref{fig:ellipsis} 
shows the two-dimensional 95\% confidence region for limits including a 
form factor with $\Lambda_{\textrm{cutoff}}=$~500~GeV.

\section{Summary}
\label{sec:summary}


Results on the search for the exclusive two-photon $W$ pair production in the 
$e^{\pm}\mu^{\mp}$ decay channel are reported. The studies are 
performed using 5.05/fb of data collected by the CMS detector in 2011 for 
proton-proton collisions at $\sqrt{s}=$~7~TeV. Two events are observed in data, 
which are in agreement with the theoretical prediction of 2.2~$\pm$~0.4 signal 
events and 0.84~$\pm$~0.15 background events, obtained in the region of
$|\eta(\ell)|<$~2.4, $p_{T}(\ell)>$~20~GeV and $p_{T}(e^{\pm}\mu^{\mp})>$~30~GeV.

Besides, further studies regarding the anomalous quartic gauge couplings are 
performed for the region with $p_{T}(e^{\pm}\mu^{\mp})>$~100~GeV. No events 
are observed in data and limits on the anomalous parameters are addresses, which 
are two orders of magnitude more stringent than those obtained at LEP and 20 
times more stringent than the results obtained at the Tevatron. Reclent results
from CMS on the tri-boson production show similar results for these limits.


\section{Acknowledgements}
\label{sec:acknow}


GGS acknowledges the support by CNPq/Brazil.


\bibliography{DASILVEIRA_GustavoGil}

\begin{thebibliography}{16}

\bibitem{Chatrchyan:2011ci}
S.~Chatrchyan et~al. (CMS), JHEP \textbf{01}, 052 (2012), \texttt{1111.5536}

\bibitem{Abazov:2013opa}
V.M. Abazov et~al. (D0 Collaboration), Phys.Rev. \textbf{D88}, 012005 (2013),
  \texttt{1305.1258}

\bibitem{Chatrchyan:2013foa}
S.~Chatrchyan et~al. (CMS Collaboration), JHEP \textbf{1307}, 116 (2013),
  \texttt{1305.5596}

\bibitem{Belanger:1992qh}
G.~Belanger, F.~Boudjema, Phys. Lett. B \textbf{288}, 201 (1992)

\bibitem{Goncalves:2007vi}
V.~Gon{\c{c}}alves, M.~Machado, Eur. Phys. J. C \textbf{56}, 33 (2008),
  \texttt{0710.4287}

\bibitem{Motyka:2008ac}
L.~Motyka, G.~Watt, Phys. Rev. D \textbf{78}, 014023 (2008), \texttt{0805.2113}

\bibitem{Cisek:2009hp}
A.~Cisek, W.~Schafer, A.~Szczurek, Phys. Rev. D \textbf{80}, 074013 (2009),
  \texttt{0906.1739}

\bibitem{Baranov:1991yq}
S.~Baranov, O.~Duenger, H.~Shooshtari, J.~Vermaseren, \emph{{LPAIR}: A
  generator for lepton pair production}, in \emph{Hamburg 1991, Proceedings,
  Physics at HERA} (1991), Vol.~3, p. 1478

\bibitem{Vermaseren:1982cz}
J.~Vermaseren, Nucl. Phys. B \textbf{229}, 347 (1983)

\bibitem{Feldman:1997qc}
G.J. Feldman, R.D. Cousins, Phys.Rev. \textbf{D57}, 3873 (1998),
  \texttt{physics/9711021}

\bibitem{Chapon:2009hh}
E.~Chapon, C.~Royon, O.~Kepka, Phys. Rev. D \textbf{81}, 074003 (2010),
  \texttt{0912.5161}

\bibitem{Achard:2001eg}
P.~Achard et~al. (L3), Phys. Lett. B \textbf{527}, 29 (2002),
  \texttt{hep-ex/0111029}

\bibitem{Achard:2002iz}
P.~Achard et~al. (L3), Phys. Lett. B \textbf{540}, 43 (2002),
  \texttt{hep-ex/0206050}

\bibitem{Abdallah:2003xn}
J.~Abdallah et~al. (DELPHI), Eur. Phys. J. C \textbf{31}, 139 (2003),
  \texttt{hep-ex/0311004}

\bibitem{Abbiendi:1999aa}
G.~Abbiendi et~al. (OPAL), Phys. Lett. B \textbf{471}, 293 (1999),
  \texttt{hep-ex/9910069}

\bibitem{CMS:2013kea}
{CMS Collaboration} (2013), \texttt{CMS-PAS-SMP-13-009}

\end{thebibliography}


\end{document}